\documentclass[journal]{IEEEtran}

\ifCLASSINFOpdf
\else
   \usepackage[dvips]{graphicx}
\fi
\usepackage{url}

\usepackage{graphicx}
\usepackage{amsmath,bm,graphicx}
\usepackage{amsfonts}
\usepackage{multirow}

\begin{document}

\title{Long-frame-shift Neural Speech Phase Prediction with Spectral Continuity Enhancement and Interpolation Error Compensation}

\author{Yang~Ai,~\IEEEmembership{Member,~IEEE},Ye-Xin Lu,~Zhen-Hua~Ling,~\IEEEmembership{Senior Member,~IEEE}
\thanks{This work was supported by the Fundamental Research Funds for the Central Universities under Grant WK2100000033. (Corresponding author: Zhen-Hua Ling)}
\thanks{Yang Ai, Ye-Xin Lu and Zhen-Hua Ling are with the National Engineering Research Center of Speech and Language Information Processing, University of Science and Technology of China, Hefei, 230027, China (e-mail: yangai@ustc.edu.cn, yxlu0102@mail.ustc.edu.cn, zhling@ustc.edu.cn).}}

\maketitle

\begin{abstract}
Speech phase prediction, which is a significant research focus in the field of signal processing, aims to recover speech phase spectra from amplitude-related features.
However, existing speech phase prediction methods are constrained to recovering phase spectra with short frame shifts, which are considerably smaller than the theoretical upper bound required for exact waveform reconstruction of short-time Fourier transform (STFT).
To tackle this issue, we present a novel long-frame-shift neural speech phase prediction (LFS-NSPP) method which enables precise prediction of long-frame-shift phase spectra from long-frame-shift log amplitude spectra.
The proposed method consists of three stages: interpolation, prediction and decimation.
The short-frame-shift log amplitude spectra are first constructed from long-frame-shift ones through frequency-by-frequency interpolation to enhance the spectral continuity, and then employed to predict short-frame-shift phase spectra using an NSPP model, thereby compensating for interpolation errors.
Ultimately, the long-frame-shift phase spectra are obtained from short-frame-shift ones through frame-by-frame decimation.
Experimental results show that the proposed LFS-NSPP method can yield superior quality in predicting long-frame-shift phase spectra than the original NSPP model and other signal-processing-based phase estimation algorithms.
\end{abstract}

\begin{IEEEkeywords}
speech phase prediction, interpolation, spectral continuity, error compensation, neural network
\end{IEEEkeywords}

\IEEEpeerreviewmaketitle

\section{Introduction}

\IEEEPARstart{S}{peech} phase prediction aims to recover phase spectra from input amplitude-related features, which is a crucial final step for reconstructing speech waveforms in many speech generation methods [1][2][3].
Due to the wrapping and distribution unstructured characteristics of phases, speech phase prediction has always been a difficult problem in the field of speech signal processing.
In the early days, researchers mainly focused on signal processing (SP) based phase estimation algorithms, such as Griffin-Lim algorithm (GLA) [4], fast GLA [5], alternating direction method of multipliers (ADMM) [6] and relaxed averaged alternating reflections (RAAR) [7], which iteratively estimated phase spectra from amplitude spectra based on alternating projection or reflection.
Recently, neural network (NN) based phase prediction methods have emerged gradually, mainly including GLA simulation methods [8][9][10], two-stage methods [11][12][13] and prior-distribution-aware methods [14][15].
However, SP-based algorithms exhibited problems such as low efficiency and poor robustness, while NN-based methods still required the assistance of conventional algorithms, resulting in complex operational procedures.

\begin{figure}
\vspace{-1mm}
    \centering
    \includegraphics[height=4.8cm]{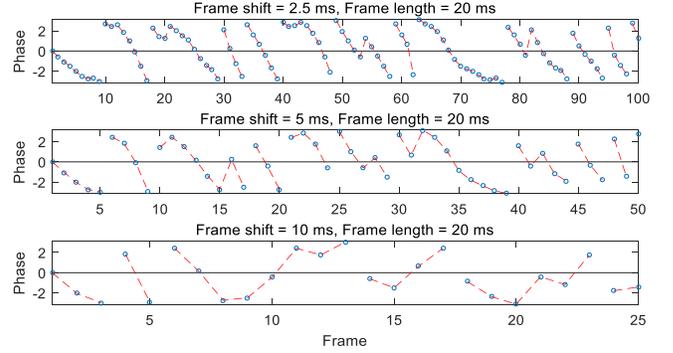}
    \caption{Scatter diagrams of the phase varying with frames at a certain frequency bin of a 0.25s 16kHz speech, when using different frame shifts (2.5 ms, 5 ms and 10 ms) and the same frame length (20 ms). Here, blue circles and red dotted lines represent phase values and envelopes, respectively.
    }
    \label{fig: Phase_continuity}
    \vspace{-3mm}
\end{figure}

To overcome the issues existed in SP-based algorithms and NN-based methods, we have proposed a neural speech phase prediction (NSPP) model [16] which can directly predict wrapped phase spectra from input log amplitude spectra (LAS).
The NSPP model is a cascade of a residual convolutional network (RCNet) and a parallel estimation architecture.
The parallel estimation architecture is one of the core components for successful phase prediction, as it guarantees direct prediction of wrapped phase spectra by neural networks through parallel pseudo real and imaginary part prediction structure and phase calculation formula.
Another core of the NSPP is the anti-wrapping losses.
They adopt a linear anti-wrapping function to activate the direct error between predicted phase spectra and natural ones, which can avoid the error expansion issues caused by phase wrapping and enable precise phase prediction.
For more details, please refer to our previous publication [16].

\begin{figure*}
    \centering
    \includegraphics[height=4.2cm]{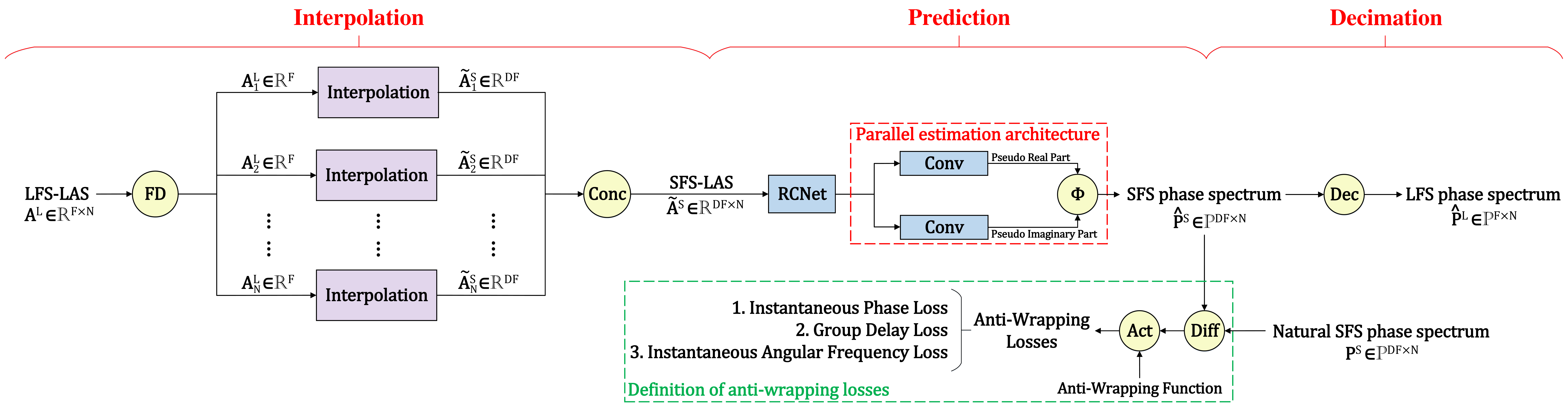}
    \caption{Overall architecture of the proposed LFS-NSPP method. Here, \emph{FD}, \emph{Conc}, \emph{$\Phi$}, \emph{Dec}, \emph{Diff} and \emph{Act} represent frequency division, concatenation, phase calculation formula, decimation, differential and activation, respectively. The blue boxes represent the trainable modules.
    }
    \label{fig: overall_architecture}
\end{figure*}

The theoretical upper bound (TUB) frame shift for exact waveform reconstruction of short-time Fourier transform (STFT) is approximately half of the frame length [17].
However, many phase prediction methods usually employ an STFT frame shift of one-eighth or one-quarter of the frame length, which is much smaller than the TUB [18].
Figure \ref{fig: Phase_continuity} provides scatter diagrams showing how the phase varies with frames when using different frame shifts as an example.
Obviously, as the frame shift increases, the temporal continuity of the phase progressively deteriorates, leading to a weakening of the inter-frame correlation.
Due to the temporal discontinuity of long-frame-shift (LFS) phase spectra, current methods may face challenges in accurately modeling it, leading to a significant decline in the precision of phase predictions.
The prediction of LFS spectra is essential in various speech generation tasks.
For instance, generating short-frame-shift (SFS) features using attention-based acoustic models [19][20] can be challenging and fail to properly align texts and features.
Additionally, predicting LFS features can also enhance the efficiency of model training and generation compared to SFS features.
Hence, predicting LFS phase spectra holds significant importance.

Therefore, this paper proposes a novel LFS-NSPP method that achieves precise prediction of LFS phase spectra from input LFS-LAS, integrating interpolation and decimation techniques built upon the foundation of NSPP [16].
Compared with the NSPP [16], the contribution of the LFS-NSPP lies in the implementation of phase prediction with long frame shifts.
The LFS-NSPP is a three-stage method, including interpolation, prediction and decimation.
At the interpolation stage, the SFS-LAS are created by interpolating input LFS-LAS in a frequency-by-frequency manner to enhance the spectral continuity.
At the prediction stage, the NSPP model leverages interpolated SFS-LAS to predict SFS phase spectra and effectively compensates for interpolation errors.
Ultimately, the LFS phase spectra are obtained through a frame-by-frame decimation of the predicted SFS phase spectra during the decimation stage.
Experimental results demonstrate that the proposed LFS-NSPP method outperforms conventional GLA and RAAR algorithms and original NSPP model in terms of subjective evaluations, yielding higher quality of reconstructed LFS phase spectra.

\begin{figure}
\vspace{-3mm}
    \centering
    \includegraphics[height=1.6cm]{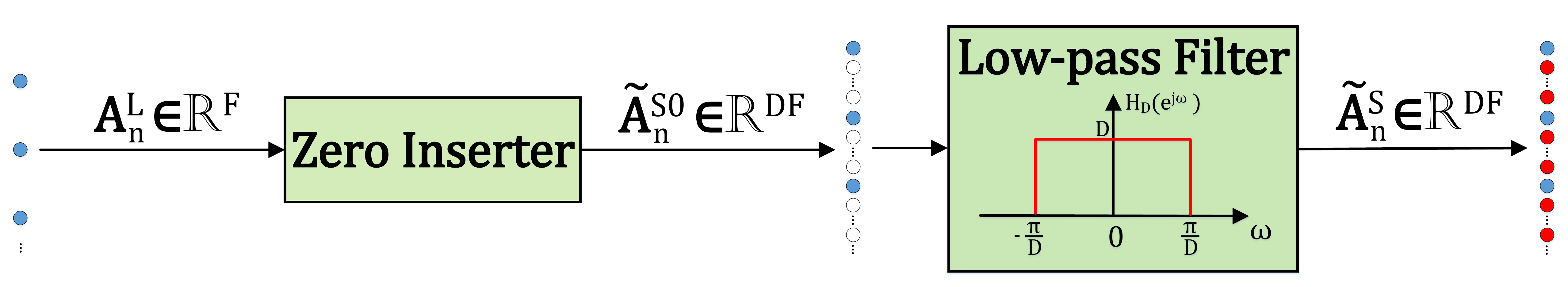}
    \caption{An illustration of the interpolation module.
    }
    \label{fig: lossless_interp}
    \vspace{-3mm}
\end{figure}

\section{Proposed Methods}

To address the challenge of predicting LFS phase spectra, this paper proposes a solution called LFS-NSPP.
As shown in Figure \ref{fig: overall_architecture}, the proposed LFS-NSPP consists of three stages: interpolation, prediction and decimation.
Overall, the LFS-NSPP uses LFS-LAS $\bm{A}^L \in \mathbb{R}^{F\times N}$ with a frame shift of $T/F$ as input and outputs LFS phase spectrum $\hat{\bm{P}}^L \in \mathbb{R}^{F\times N}$ with the same frame shift. $T$ represents the number of time-domain waveform points, and $F$ and $N$ respectively represent the number of spectral frames and frequency bins.

Theoretically, the interpolation should be performed on the time series of the LFS complex spectrum, and then the SFS-LAS is extracted from it and fed into the prediction stage.
This interpolation process is lossless.
However, due to the unknown phase information, we adopt the method of LAS interpolation, which assumes the phase is zero.
Therefore, the interpolation method we use introduces an unavoidable \emph{interpolation error} $e_m$, compared to the interpolation result of the complex spectrum, i.e.,
\begin{equation}
\label{equ: d}
e_i=\left\|I\left(\bm{A}_n^L\right)-\log\left|I\left(e^{\bm{A}_n^L}e^{j\bm{P}_n^L}\right)\right| \right\|_1>0,
\end{equation}
where $\bm{A}_n^L=\bm{A}^L(:,n)$ and $\bm{P}_n^L=\bm{P}^L(:,n)$ are the natural LFS-LAS and LFS phase spectrum at the $n$-th frequency bin ($n=1,\dots,N$), respectively.
$I(\cdot)$ denotes the interpolation operation.
As a solution, during the training process, we utilize the interpolated SFS-LAS as the input to the NSPP model at the prediction stage rather than using the natural SFS-LAS.
This approach is undertaken with the expectation that the NSPP model can acquire the ability to compensate for the interpolation error through the training procedure.

Another approach to address this issue is the iterative LFS-NSPP.
The method involves multiple iterations of using LFS-NSPP to predict the phase and continuously updating the phase information in the complex spectrum interpolation.
However, this method inevitably leads to a decrease in generation efficiency, which is not considered in this paper.

\vspace{-1mm}
\subsection{Interpolation}
\vspace{-0.1mm}

During the interpolation stage, in order to enhance the spectral continuity, the LFS-LAS $\bm{A}^L \in \mathbb{R}^{F\times N}$ uses interpolation technique to fill in and create SFS-LAS $\tilde{\bm{A}}^S \in \mathbb{R}^{DF\times N}$ with a frame shift of $T/(DF)$, where $D$ is the ratio between the long frame shift and the short frame shift.
The number of frames for interpolated $\tilde{\bm{A}}^S$ is $D$ times that of $\bm{A}^L$.

Specifically, as shown in Figure \ref{fig: overall_architecture}, the LFS-LAS $\bm{A}^L$ firstly performs frequency division to obtain $N$ sequences $\bm{A}_n^{L},n=1,\dots,N$, each corresponding to a specific frequency bin.
Secondly, each sequence passes through the same interpolation module, which inserts $D-1$ additional samples after each existing sample in the sequence to construct a new interpolated sequence $\tilde{\bm{A}}_n^{S} \in \mathbb{R}^{DF}$ whose length is $D$ times that of the original one.
The interpolation operation enhances the spectral temporal continuity while maintaining the original spectral information.
Finally, these interpolated sequences are concatenated along the frequency axis to construct the SFS-LAS, i.e.,
\begin{equation}
\label{equ: Conc}
\tilde{\bm{A}}^S=\left[\tilde{\bm{A}}_1^{S},\tilde{\bm{A}}_2^{S},\dots,\tilde{\bm{A}}_N^{S} \right] \in \mathbb{R}^{DF\times N}.
\end{equation}

As illustrated in Figure \ref{fig: lossless_interp}, the interpolation module consists of a zero inserter and a low-pass filter.
To interpolate the original sequence $\bm{A}_n^{L}$, it is first passed through a zero inserter to insert $D-1$ zeros after each sample.
This results in a sequence $\tilde{\bm{A}}_n^{S0}\in \mathbb{R}^{DF}$ of length $DF$, i.e.,
\begin{equation}
\label{equ: Zero_inserter}
\tilde{\bm{A}}_n^{S0}(f)=\left\{\begin{array}{rl}\!\bm{A}_n^{L}\left( \frac{f+D-1}{D} \right),& \frac{f+D-1}{D}\in \mathbb{N}_{+}^L \\ 0,&\frac{f+D-1}{D}\notin \mathbb{N}_{+}^L\end{array}\right.,
\end{equation}
where $f=1,2,\dots,DF$ and $\mathbb{N}_{+}^L=\{1,2,\dots,F \}$.
Subsequently, the zero-inserted sequence $\tilde{\bm{A}}_n^{S0}$ is filtered by a low-pass filter with frequency response
\begin{equation}
\label{equ: Frequency response}
\bm{H}(e^{j\omega})=\left\{\begin{array}{rl}D,& |\omega|\leq \frac{\pi}{D} \\ 0,& \frac{\pi}{D}\leq|\omega|\leq\pi \end{array}\right.,
\end{equation}
so the zero samples are replaced with interpolation samples to generate $\tilde{\bm{A}}_n^{S}$.

\vspace{-0.5mm}
\subsection{Prediction}
\vspace{-1mm}

During the prediction stage, in order to compensate for interpolation errors, the NSPP model predicts the SFS phase spectrum $\hat{\bm{P}}^S \in \mathbb{R}^{DF\times N}$ with an frame shift of $T/(DF)$ from the input interpolated SFS-LAS $\tilde{\bm{A}}^S \in \mathbb{R}^{DF\times N}$, i.e.,
\begin{equation}
\label{equ: NSPP}
\hat{\bm{P}}^S=NSPP(\tilde{\bm{A}}^S).
\end{equation}
Three anti-wrapping losses used in our previous publication [16] are also defined here.
The instantaneous phase (IP) loss, group delay (GD) loss and instantaneous angular frequency (IAF) loss defined between the predicted SFS phase spectrum $\hat{\bm{P}}^S$ and natural one $\bm{P}^S$ are used to train the trainable modules.

\vspace{-0.5mm}
\subsection{Decimation}
\vspace{-1mm}

Decimation is the final step of our proposed LFS-NSPP.
The predicted SFS phase spectrum $\hat{\bm{P}}^S \in \mathbb{R}^{DF\times N}$ decimates frames at corresponding LFS positions and constructs the LFS phase spectrum $\hat{\bm{P}}^L \in \mathbb{R}^{F\times N}$ by concatenation as follows:
\begin{equation}
\label{equ: Decimation}
\hat{\bm{P}}^L=\left[\hat{\bm{P}}_1^S,\hat{\bm{P}}_{1+D}^S,\hat{\bm{P}}_{1+2D}^S,\dots,\hat{\bm{P}}_{1+(F-1)D}^S  \right].
\end{equation}

%
%

\section{Experiments}
\label{sec: Experiments}

\subsection{Data and feature configuration}
\label{subsec: Data and feature configuration}

The VCTK corpus [21] was adopted in our experiments\footnote{Source codes are available at \url{https://github.com/yangai520/LFS-NSPP}. Examples of generated speech can be found at \url{https://yangai520.github.io/LFS-NSPP}.}.
We selected 43,411 utterances from 107 speakers and randomly divided them into a training set (42,851 utterances) and a validation set (560 utterances).
We then built the test set, which included 824 utterances from 2 unseen speakers (a male speaker and a female speaker).
The original waveforms were downsampled to 16 kHz for the experiments.
The Hanning window was used in STFT.
The frame length of the LFS-LAS and LFS phase spectra was set to 20 ms.
The frame shift was set to 10 ms (i.e., $T/F=160$) which reached the TUB.
The FFT point number was 1024 (i.e., $N=513$).
The configuration of the RCNet and parallel estimation architecture in the LFS-NSPP was borrowed from the original NSPP [16].

\subsection{Effects of frame shift on phase prediction methods}
\label{subsec: Effects of frame shift on phase prediction methods}


\begin{table}
\centering
    \caption{MOS score and RTF results of GLA, RAAR and original NSPP for frame shifts of three different lengths. ``$a\times$" represents $a\times$ real time.}
    \begin{tabular}{c | c c | c | c}
        \hline
        \hline
         \textbf{Frame shift} & \textbf{Amplitude} & \textbf{Phase} & \textbf{MOS}$\uparrow$ & \textbf{RTF}$\downarrow$\\
         \hline
         --& Natural & Natural & 4.08$\pm$0.070 & --\\
         \hline
         \multirow{3}{*}{2.5 ms}
         & Natural & GLA & 3.90$\pm$0.078 & 0.58 (1.72$\times$) \\
         & Natural & RAAR & \textbf{4.02$\pm$0.073} & 0.72 (1.38$\times$) \\
         & Natural & NSPP & \textbf{4.01$\pm$0.070} &  \textbf{0.10 (10.0$\times$)}\\
         \cline{1-5}
         \multirow{3}{*}{5 ms}
         & Natural & GLA & 3.77$\pm$0.079 & 0.32 (3.13$\times$) \\
         & Natural & RAAR & \textbf{3.99$\pm$0.078} & 0.44 (2.27$\times$) \\
         & Natural & NSPP & \textbf{3.97$\pm$0.076} & \textbf{0.063 (15.9$\times$)} \\
         \cline{1-5}
         \multirow{3}{*}{10 ms}
         & Natural & GLA & 2.20$\pm$0.079 &  0.30 (3.33$\times$)\\
         & Natural & RAAR & 2.68$\pm$0.11 &  0.46 (2.17$\times$)\\
         & Natural & NSPP & \textbf{2.75$\pm$0.091} & \textbf{0.041 (24.4$\times$)}\\
         \cline{1-5}
        \hline
        \hline
    \end{tabular}
\label{tab: FS-MOS}
\end{table}

We first validated the effect of frame shift on existing phase prediction methods.
Mean opinion score (MOS) tests [22][23] were conducted to compare the naturalness of speeches reconstructed by GLA [4], RAAR [7] and original NSPP [16] for frame shifts of three different lengths (i.e., 2.5 ms, 5 ms and 10 ms).
In each MOS test, 20 test utterances reconstructed by these methods along with the natural utterances were evaluated by at least 30 native English listeners on the crowdsourcing platform of Amazon Mechanical Turk (AMT).
Listeners were asked to give a naturalness score between 1 and 5, and the score interval was set to 0.5 to better assess the differences among systems.
To evaluate the generation efficiency, the real-time factor (RTF), which is defined as the ratio between the time consumed to generate all test sentences using a single Intel Xeon E5-2680 CPU core and the total duration of the test set, was also utilized as an objective metric.

The MOS score and RTF results are shown in Table \ref{tab: FS-MOS}.
The commonality among the three phase prediction methods is that the MOS score decreased as the frame shift increased.
The speech quality experienced a significant decline when the frame shift was increased from 5 ms to 10 ms (i.e., the TUB), indicating that the available phase prediction methods were inadequate for predicting the LFS phase spectra.
Considering both speech quality and generation efficiency, our originally proposed NSPP demonstrated remarkable advantages over the other two SP-based algorithms.
Therefore, we have selected NSPP as the fundamental basis for this study.

\begin{table}
\centering
    \caption{Instantaneous angular frequency (IAF) losses of LFS phase spectra predicted by NSPP-based methods on the test set.}
    \begin{tabular}{c | c c c c}
        \hline
        \hline
        & \textbf{LFS-NSPP} & \textbf{NSPP} & \textbf{LFS-NSPP-D4} & \textbf{LFS-NSPP*}\\
         \hline
         IAF & \textbf{1.19} & 1.35 & 1.22 & 1.33\\

        \hline
        \hline
    \end{tabular}
\label{tab: Objective}
\end{table}

\begin{figure}
    \centering
    \includegraphics[height=2.9cm]{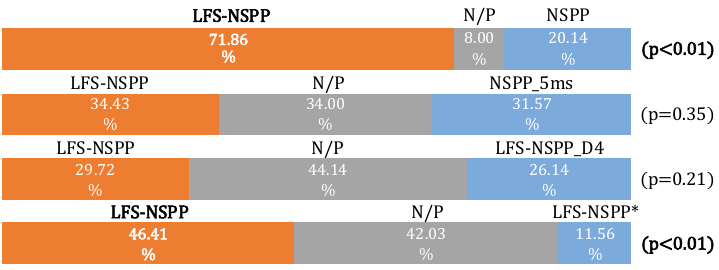}
    \caption{Average preference scores (\%) of ABX tests on speech quality for comparison among NSPP-based phase prediction methods, where N/P stands for ``no preference" and $p$ denotes the $p$-value of a $t$-test between two methods.
    }
    \label{fig: ABX1}
\end{figure}

\subsection{Comparison among NSPP-based phase prediction methods}
\label{subsec: Comparison among NSPP-based methods}

We subsequently carried out four groups of ABX preference experiments on AMT to compare the performance of various NSPP-based methods in predicting LFS phase spectra with frame shift of 10 ms.
The listeners were asked to judge which utterance in each paired speeches had better quality or whether there was no preference.
The ABX test results are shown in Figure \ref{fig: ABX1}.
These methods are introduced as follows:
\begin{itemize}
\item {}{\textbf{LFS-NSPP}}: The proposed LFS-NSPP method with the interpolation ratio of $D=2$.

\item {}{\textbf{NSPP}}: The original NSPP model [16] trained using natural LFS-LAS with frame shift of 10 ms.

\item {}{\textbf{NSPP-5ms}}: The original NSPP model [16] trained using natural SFS-LAS with frame shift of 5 ms.

\item {}{\textbf{LFS-NSPP-D4}}: The proposed LFS-NSPP method with the interpolation ratio of $D=4$.

\item {}{\textbf{LFS-NSPP*}}: Compared with \textbf{LFS-NSPP}, the NSPP model at the prediction stage was directly replaced with \textbf{NSPP-5ms}.

\end{itemize}

\begin{figure}
    \centering
    \includegraphics[height=4.5cm]{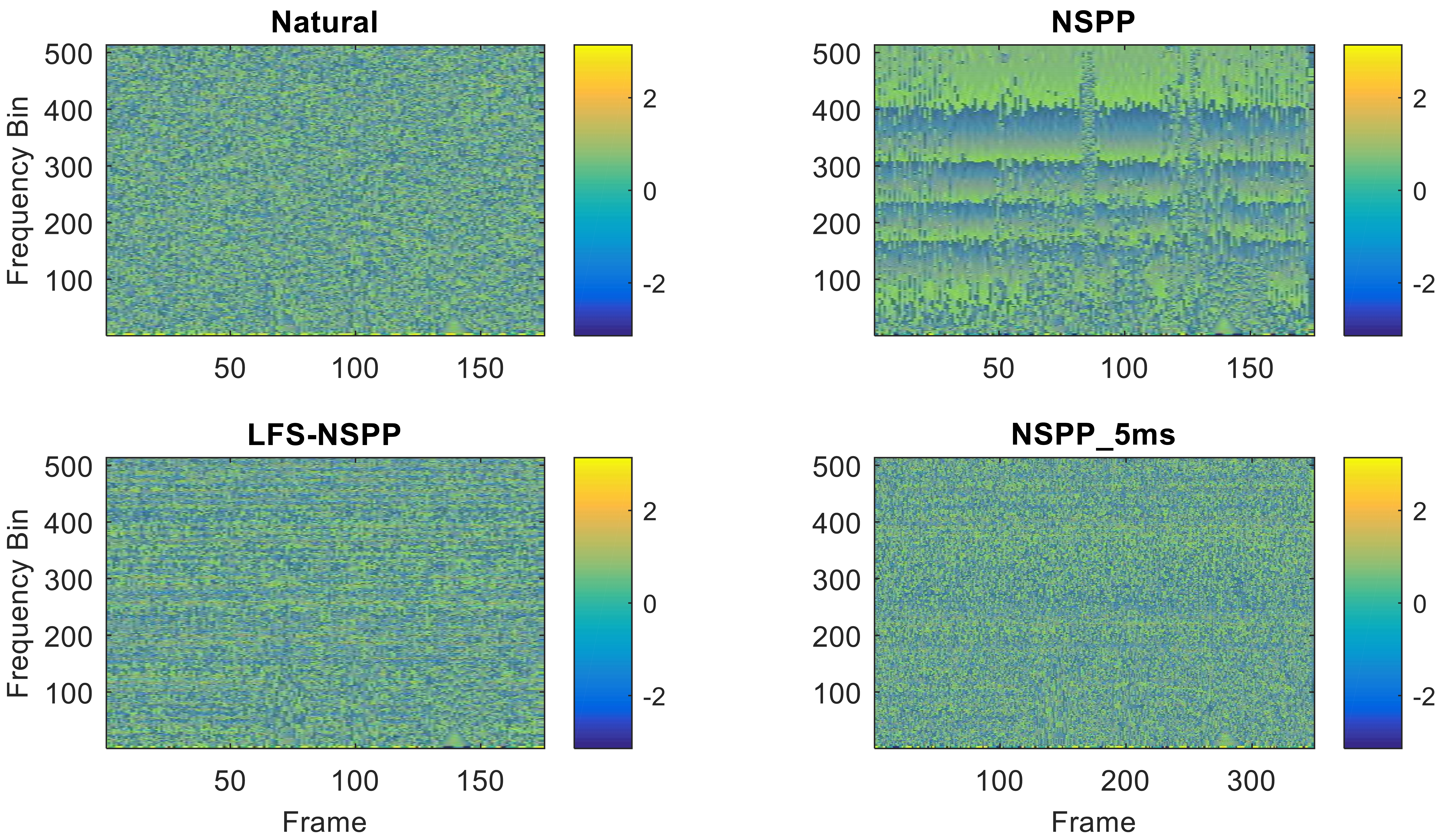}
    \caption{Color maps of natural phase spectrum (10 ms frame shift) and phase spectra predicted by \textbf{NSPP}, \textbf{LFS-NSPP} and \textbf{NSPP\_5ms} for a test utterance.
    }
    \label{fig: Color_map}
\end{figure}

Obviously, the proposed \textbf{LFS-NSPP} outperformed the \textbf{NSPP} significantly ($p <$ 0.01).
As shown in Figure \ref{fig: Color_map}, the phase spectrum produced by \textbf{NSPP} exhibited noticeable horizontal lines especially in the high-frequency range, resulting in an unpleasant buzzing noise.
We also calculated the IAF loss used in NSPP [16] of the LFS phase spectra generated by \textbf{LFS-NSPP} and \textbf{NSPP} on the test set for objective evaluations.
This loss captures the actual gap between the temporal difference of the predicted phase spectra and natural ones.
Table \ref{tab: Objective} reveals that the IAF loss decreased from 1.35 to 1.19 when transitioning from \textbf{NSPP} to \textbf{LFS-NSPP}, confirming an improvement in the temporal continuity of phase spectra.
Moreover, the \textbf{LFS-NSPP} was also competitive with \textbf{NSPP-5ms} ($p >$ 0.05) by observing Figure \ref{fig: ABX1} and \ref{fig: Color_map}.
All of these results demonstrated the efficacy of the LFS-NSPP in predicting LFS phase spectra whose frame shift reached the TUB.
Furthermore, we evaluated the performance of LFS-NSPP under various interpolation rates.
There was no significant difference ($p >$ 0.05) between the subjective quality and IAF results of \textbf{LFS-NSPP} and \textbf{LFS-NSPP-D4} as shown in Figure \ref{fig: ABX1} and Table \ref{tab: Objective}, which indicated that our proposed method will work as long as the spectral continuity is significantly improved, regardless of the interpolation rate used.
Figure \ref{fig: ABX1} and Table \ref{tab: Objective} also reveal that the \textbf{LFS-NSPP} outperformed \textbf{LFS-NSPP*} significantly ($p <$ 0.01), confirming that the proposed method can compensate for interpolation errors.

\begin{figure}
    \centering
    \includegraphics[height=2.9cm]{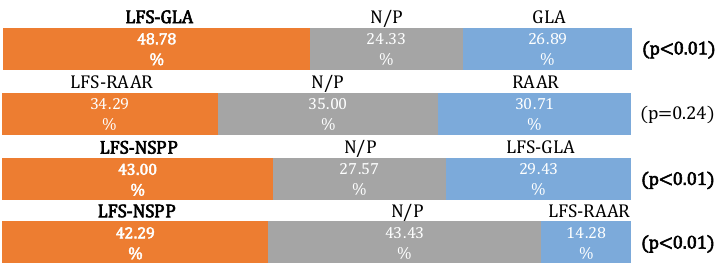}
    \caption{Average preference scores (\%) of ABX tests on speech quality for comparison with SP-based phase estimation algorithms, where N/P stands for ``no preference" and $p$ denotes the $p$-value of a $t$-test between two methods.
    }
    \label{fig: ABX2}
\end{figure}

\subsection{Comparison with SP-based phase estimation algorithms}
\label{subsec: Comparison with SP-based phase estimation algorithms}

Finally, we compared the proposed LFS-NSPP (i.e., the \textbf{LFS-NSPP} used in Section \ref{subsec: Comparison among NSPP-based methods}) with GLA and RAAR by ABX tests.
For fairness, the GLA and RAAR methods with 100 iteration rounds were also applied using the proposed interpolation-decimation framework to construct LFS-GLA and LFS-RAAR, respectively.
The ABX test results in Figure \ref{fig: ABX2} demonstrated that the proposed interpolation-decimation framework also contributed to improving the performance of GLA and RAAR to some extent.
However, there was still a noticeable disparity between the performance of GLA and RAAR and the proposed LFS-NSPP.
This can be attributed to the fact that SP-based algorithms were unable to learn to compensate for interpolation errors by training.
This further validated the effectiveness of the proposed method.

\section{Conclusion}
\label{sec: Conclusion}

In this paper, we have proposed LFS-NSPP which can precisely predict the TUB-reached LFS phase spectra from LFS-LAS.
There were three stages in the generation process of the LFS-NSPP, namely interpolation, prediction and decimation.
The core of the LFS-NSPP was to enhance the spectral continuity by employing interpolation technique and compensate for interpolation errors by NSPP model training.
Experimental results show that the proposed LFS-NSPP outperformed the SP-based GLA and RAAR algorithms and original NSPP model.
Applying the LFS-NSPP to specific speech generation tasks will be the focus of our future work.


\end{document}